\documentstyle[12pt]{article}
\begin{document}

\begin{center}
               Color Transparency and Fermi Motion\\
               Byron K. Jennings,\\
               {\em TRIUMF, 4004 Wesbrook Mall, Vancouver, B.C. V6T 2A3, Canada
               }\\
               Boris Z. Kopeliovich,\\
               {\em Joint Institute of Nuclear Research,
               Head Post Office, P.O. Box 79,
               10100 Moscow,
               Russia}\\
\end{center}

\begin{center}
Abstract
\end{center}

It is argued that color transparency in quasi elastic scattering of electrons
and hadrons on nuclei is possible only due to Fermi-motion. We found a strong
dependence of nuclear transparency on Bjorken x in (e,e'p), it is close to the
Glauber model expectations at $x>1$, but increases and even exceeds one at
$x<1$. It is argued that color transparency is accompanied by large
longitudinal momentum transfer to nuclear matter during the passage of the
small size wave packet.  \vspace{\fill}

e-mail\\
boris@bethe.npl.washington.edu\\
jennings@decu07.triumf.ca
\pagebreak

The phenomenon of color transparency (CT) provides an unique test of the
dynamical role of color in QCD. The screening of the intrinsic color of
colorless hadrons leads to a weak nonexponential attenuation of high energy
hadrons in nuclear matter \cite{zamol}, filtering of transverse sizes through
nuclei \cite{bertch} and to the vanishing of initial and final state
interactions of hadrons participating in a hard reaction on nuclei
\cite{mueller,brodsky}.

The strength of attenuation of hadrons in nuclei can be judged by measuring
the A dependence of cross-sections for nuclear reactions. Data are usually
presented in the form of a ratio $Tr = \sigma^A/(A \sigma^N)$ called nuclear
transparency. Here $\sigma^{A,N}$ are cross sections of a reaction on a nucleus
and on a free nucleon respectively, with the same kinematics in each case.

Quantum-mechanical interference effects violate many of the expectations of the
classical approach. It is demonstrated in ref.~\cite{kop91a,kop91b} that
nuclear transparency, $Tr$, can exceed one. So it doesn't have a
meaning of transparency of the nucleus. It also leads to an enhancement of
polarization effects\cite{zak87}, an obvious indication of interference.
Contrary to naive expectations attenuation of photoproduction of $\psi'$
charmonium is much weaker then of $J/\psi$ which has a smaller
radius\cite{kop91a}.

The purpose of the present paper is to develop the quantum mechanical treatment
of CT on, for example, the $A(e,e'p)A'$ reaction. We find dramatic
effects. Nuclear transparency is strongly dependent on the
Bjorken variable $x_B$ or equivalently the lost momentum of the struck
nucleon. At $x_B > 1$ CT effects are negligibly small. The main effects are
localized at $x_B < 1$, and should be easily observed at $Q^2 > 10 GeV^2$.
At lower $Q^2$ CT effects are predicted to be so weak that their experimental
detection is doubtful.

According to wide spread folklore interaction of small size objects
disappears, due to the effect of color screening. However literally speaking
this is not true. We argue here that such objects strongly interact and
transfer considerable longitudinal momentum to the nuclear matter, only the net
attenuation of these objects vanishes.

CT, as first stated in the paper ref.~\cite{zamol}, is a
particular case of Gribov's inelastic shadowing \cite{gr69}. In hadronic state
representation diagonal and off diagonal diffractive amplitudes cancel and
attenuation vanishes. The possibility of such fine tuning of diffractive
amplitudes was considered earlier \cite{kop78,kop80}.  This was also realized
later in a ref.~\cite{jen90,jen91}.  Evolution can also be calculated in the
quark basis and a path integral method was developed in\cite{kop91a,kop91b}.

In this note we investigate the role of Fermi motion and longitudinal momentum
transfer in nuclear transparency. Let us consider elastic ep scattering with
high $Q^2$.  We will consider the imaginary part of the amplitude corresponding
to all intermediate particles being on mass shell.  According to expectations
of PQCD average size of the ejectile is small and it has to demonstrate CT. To
have CT different intermediate states have to cancel each other.  However at
fixed $x_B=1$ only the proton can be produced on shell if the target proton is
at rest. Thus no cancellation is possible, ie.\ no CT.  However in quantum
mechanics how the size is measured must be defined. To observe CT the size
detector (second scattering center) must be put at a short distance from the
target proton. Then the target proton can not be treated as at rest, it has
uncertainty in momentum. As a result, some heavier states can be produced. The
closer the target is to the size detector the more states are produced, the
complete is the cancellation.

Thus the Fermi motion of the bound nucleon is a source of CT in quasielastic
scattering. On the other hand it restricts the amount of CT. At $x_B = Q^2/(2
m_p \nu) =1$, to produce an excited state of mass $m^*$ in the final state, the
target nucleon must preferentially have an initial momentum in the direction
opposite to the photon, $k_p < -(m^{*2} + m_p^2)/2\nu$. So the mass spectrum of
produced states is restricted by:
\begin{equation}
m^{*2} < m_p^2 + 2 \nu k_F.
\end{equation}
Consequently even if the ejectile has a very small size,
$\rho^2 \approx 1/Q^2$, the nucleus, as a quantum size detector, is insensitive
to such small sizes. It can resolve only sizes larger than:
\begin{equation}
\rho^2 \geq \frac{m_p}{k_F}\frac{1}{Q^2}.
\end{equation}
We see that the $(e,e'p)$ and $(p,2p)$ reactions have poor kinematics for CT
studies. One has to increase $Q^2$ which decreases the cross-section while one
 just wants to increase the laboratory frame momentum of the final
particle.

Nevertheless the sensitivity of the nucleus to the size of the ejectile can be
enhanced, effectively broadening the mass spectrum of produced particles.  This
is done by decreasing $x_B$ to about $1-k_F/m_p$.  Then all Fermi
momenta are used, the range of $m^{*2}-m^2_p$ is doubled and the nucleus can
analyze $\rho^2$ a factor of two smaller.

For numerical estimates we need some reasonable approximations. We replace the
full expansion of the hadron wave function over states with a fixed
transverse size $\rho$ by the two-component approximation,
\begin{equation}
|h\rangle \approx \alpha | 0\rangle +  \sqrt{1 -\alpha^2} | 1\rangle
\label{4}
\end{equation}
This model, as well as the eigen state method was first proposed for
calculations of inelastic correction in \cite{kop78,kop80}. The two component
method has more recently been used by in ref.~\cite{jen90,jen91}.

The state $|0\rangle$ in eq.~\ref{4} effectively includes all states
$|\rho\rangle$ with small attenuation in the nucleus, ie. $\rm{Im}\;f(\rho) \ll
1/\rho_AR_A$ where  $f(\rho)$ is the eigenvalue of the scattering amplitude,
$\rho_A$ and $R_A$ are the nuclear density and radius respectively. We consider
the state $|0\rangle$ as non-interactive, ie $f_0=0$. The corresponding
effective amplitude $\rm{Im} f_1 = \sigma^{hN}_{\rm tot} /2 (1-\alpha^2)$.

The only unknown parameter, $\alpha$, can be fixed, using the forward
diffractive dissociation cross-section, $\sigma^{hN}_{\rm dd}$ and elastic
cross-section $\sigma^{hN}_{\rm el}$\cite{kop78}:
\begin{equation}
\frac{\sigma_{\rm dd}}{\sigma_{\rm el}} = \frac{\alpha^2}{1-
\alpha^2}.\label{five}
\end{equation}
Here the eigenvalues of the scattering amplitude are averaged over the
eigenstates of the interaction.  The value of the ratio in eq.~\ref{five} is
known\cite{blattel} to be about 0.1 for pp scattering, so we use
$\alpha^2=0.1$. The value of 0.5 was used in ref.~\cite{jen91}. The results
are quite insensitive to the exact value of $\alpha^2$.

The next step is to assume that only the small size state is produced in the
$(e,e'p)$ reaction. The Fermi motion provides different weight factors for
different hadronic states, depending on their mass. Instead of the pure
initial, produced on a a free proton,
$|i\rangle = |0\rangle = \alpha |p\rangle + \sqrt{1- \alpha^2}|p^*\rangle$ we
get:
\begin{equation}
|\tilde{i}\rangle = \alpha |p\rangle + \sqrt{1-\alpha}|p^*\rangle
\sqrt{\frac{W_A(x_B-\Delta x_B)}{W_A(x_B)}}.\label{six}
\end{equation}
Here $|p^*\rangle$ is a state of mass $m^*$, which takes effectively into
account all the diffractive excitation of the proton. The hadronic basis
$|p\rangle$, $|p^*\rangle$ are eigenstates of the free Hamiltonian,
$W_A(x_B)$ is the probabilty for a struck proton to carry a momentum fration,
$x_B/A$, of the nucleus monentum in the infinite momentum frame.
The heavier state $|p^*\rangle$ is produced with a shifted value
of $x_B$:
\begin{equation}
\Delta x_B = \frac{m^{*2} - m_p^2}{2 m_p \nu}.\label{seven}
\end{equation}
Thus the $|p\rangle$ and $|p^*\rangle$ get different weight factors in
eq.~\ref{six}. The dominant resonances produced in the diffractive dissociation
ate the $N^*(1440)$, the $N^*(1520)$ and the $N^*(1680)$. However the
$N^*(1440)$ is strongly suppressed in electroproduction with high $Q^2$.So we
fixed $m^*=1.6$GeV.

The next step is to study the evolution of the wave packet with initial wave
function, eq.~\ref{six}, as it propagates through the nucleus. This can be done
with the equation obtained in \cite{kop80}:
\begin{equation}
i\frac{d}{dz} | P \rangle = \hat U | P \rangle\label{eight}
\end{equation}
where $|P\rangle$ denotes the set of states $|p\rangle$ and $|p^*\rangle$. The
evolution operator $\hat U$ has the form\cite{kop80}:
\begin{equation}
\hat U = \left(
  \begin{array}{cc}
    p -i \sigma^{pp}_{\rm tot}/2\;\rho_A(z) & \frac{\alpha}{\sqrt{1-
\alpha^2}} i\sigma^{pp}_{\rm tot} \rho_A(z)\\  \frac{\alpha}{\sqrt{1-
\alpha^2}} i\sigma^{pp}_{\rm tot} \rho_A(z) & p -\Delta p - i\sigma^{pp}_{\rm
tot}/2\;\rho_A(z)
  \end{array}\right)
\end{equation}
Here $z$ is the longitudinal coordinate along the proton trajectory, $p$ is the
proton momentum and $\Delta p = \Delta x_B m_p$.

We performed numerical calculations for the $^{56}$Fe nucleus with the Woods
Saxon density distribution\cite{yager}. The Bjorken variable $x_B$ is related
to the Fermi momentum of the proton by:
\begin{equation}
x_B \approx 1 - \frac{k_z}{m_p}
\end{equation}
where the positive direction of $k_z$ corresponds with the direction of the
photon. This expression is valid at small  $|k| \leq k_F$, where $k_F$ is the
average Fermi momentum. We have used a Gaussian parameterization of the Fermi
momentum distribution:
\begin{equation}
 W_A(k) = \frac{3}{2\pi k_F^2}\exp(-3k^2/2k_F^2).
\end{equation}

We have calculated the nuclear transparency as a function of $x_B$ at fixed
$Q^2$. It is defined as
\begin{equation}
Tr=
\frac
{\Sigma_{\alpha}\int
d^2b\int^{\infty}_{-\infty}dz_1\;e^{ik_zz_1}\psi^{\alpha}_A(b,z_1)
\langle p|\hat V(z_1,\infty)|i\rangle
\int^{\infty}_{-\infty}dz_2\;e^{-ik_zz_2}\psi^{\alpha}_A(b,z_2)^*
\langle p|\hat V(z_2,\infty)|i\rangle^*}
{W_A(k_z)}
\end{equation}
where $k$ is the missing momentum in the $(e,e'p)$ reaction and $\psi_A^\alpha$
is the nuclear wave function in the shell $\alpha$. $\hat V(z,z')|i\rangle
exp(ik_z(z-z'))$ is a solution of eq.~(7)
at the point $z'$ with the initial state $|i\rangle$ at the point $z$.
$W_A(k_z)=\int d^2k_T~W_A(\vec k)$. The nuclear density matrix $\rho(\vec
r_1,\vec r_2)= \sum_\alpha\psi_\alpha(\vec r_1)\psi^*(\vec r_2)$ is connectd
with the Fermi momentum distribution by a Fourier transformation: $W_A(k)=\int
d^3r_1d^3r_2 \exp(i \vec k \cdot (\vec r_1 -\vec r_2)) \rho(\vec r_1,\vec
r_2)$, so the correlation
length of $\rho(r_1,r_2)$, $\langle(\vec r_1 -\vec r_2)^2\rangle^{1/2} \approx
\frac{\sqrt 6}{ k_F}\approx 2fm$.
This correlation length satisfies the conditions,
$m_p~\Delta z\Delta x_B/2\ll 1$ at energies above a few GeV, and
$\sigma_{tot}\rho_A(\vec r)\Delta z/4\ll 1$. Therefore eq. (11) can be
modified in the first order on $\Delta z$ in the form,
\begin{equation}
Tr(x_B) = \frac{\int d^2b\;\int_{-\infty}^{\infty}dz\;\rho_A(b,z)\;
|\langle p|\hat V(z,\infty)|\tilde i\rangle |^2}{A\;|\langle
p|i\rangle|^2}
\nonumber\\
\end{equation}

    The evolution is calculated along the $z$-axis starting from the point
    $(b,z)$. The results are shown in fig.1 as a function of $x_B$ for $Q^2$
    equal to 7 GeV$^2$, 15 GeV$^2$ and 30 GeV$^2$. At $x_B > 1$ the nuclear
    transparency is small and close to the expectation of the Glauber model.
    The reason is obvious: at $x_B > 1$ the quasi elastic scattering takes
    place on protons having large negative Fermi momenta. To make the nucleus
    transparent excited states must be produced as well. The latter however
    prefer higher Fermi momenta which are suppressed. So at large $x_B$
    predominately protons are produced.

At $x_B < 1$, on the contrary excited states are preferentially produced; to
the extent that the transparency even becomes larger then 1 for small $x_B$.
This is understood as follows: at small $x_B < 1$, or equivalently large
positive $k_z > k_F$, direct proton production is strongly suppressed by the
nuclear wave function. At the same time the production of higher states needs
$k_z$ smaller; they are suppressed much less or even enhanced. The excited
state can convert to a proton during its propagation through the nucleus. As a
results the value of Tr$(x_B)$ can be larger then one.

A measurement of the $x_B$ dependence of nuclear transparency would be the best
test of CT, however that might be experimentally difficult. Instead we can
simply divide all events into two samples; one with $x_B > 1$ and the other
with $x_B< 1$. The CT effects are expected to be quite different in the these
two samples with a much larger effect seen for $x_B< 1$. Results of such a
calculation are shown in fig. 2 as a function of $Q^2$. The curve corresponding
to $x_B > 1$ does not appreciably deviate from the Glauber approximation. The
curve corresponding to $x_B< 1$, on the other hand increases steeply as a
function of $Q^2$.

Let us briefly list the main observations of this paper:

i) CT is a results of strong cancellations of the amplitudes for the production
of the proton and the production of excited states. This cancellation leads to
a high nuclear transparency, but does not mean the absence of interaction.
Production of different intermediate states use different Fermi momenta of the
target proton and different momenta returned to the nucleus during evolution of
these states through the nucleus. So naive expectations that CT, as a
suppression of final(initial) interactions, provides a good method to measure
Fermi momentum distribution is wrong --- the amount of transparency depends
strongly on the Fermi momentum of the struck particle. There is an, in
principle, uncertainty in the initial momentum of struck proton of the order of
$(m^{*2} - m_p^2)/2\nu$, which is essential even at high energies, as the
effective value of $m^*$ increases with energy.

ii) The Fermi momentum distribution is essential for the phenomenon of CT, it
gives the possibility of simultaneous production of states of different masses.
The Fermi momentum distribution modifies the relative weights of different
hadronic states in the produced wave packet. As a result the fine tuning needed
for CT is destroyed. It can be restored at high energy.

iii) The distortion of the tuning of the relative contributions of different
states, imposed by the Fermi momentum distribution, depends strongly on $x_B$.
It is most important at large $x_B > 1$, where CT disappears.
The weakest distortion of the tuning takes place at $x_B \approx 1 -k_F/m_p$.
At smaller $x_B$ the tuning is also violated, but the result is opposite: the
nuclear transparency increases and exceeds one.

iv) It is very difficult to observe CT effects below $Q^2<7$ Gev$^2$. This
includes the region explored by the recent experiment at SLAC\cite{slac} and
the region that can be explored by CEBAF.  On the other hand considerable
effects are predicted at $Q^2 > 10$ GeV$^2$.

v) The calculations  here have mostly a demonstrative character. The mass
distributions are different in deep inelastic scattering and diffractive
dissociation, with different centers of gravity. The two channel models
unable to incorporate this effect. In addition $m^{*2}$ should depend on the
energy of the interactions. More realistic Fermi momentum distributions should
be used taking into account the nuclear shell distributions and non-diagonal
effects. Never-the-less the present calculation are indicative of the
results that would be obtained with a better calculation.

vi) All the above statements are qualitatively valid for the wide angle quasi
elastic $(p,2p)$ scattering. Some of the results of the experiment performed at
BNL\cite{carroll}, that seemed puzzling, now become clearer.

Measurements\cite{carroll} were done at three incident energies. The kinematics
of the experiment was then used to reconstruct the initial Fermi momentum of
the target proton (neglecting binding energy and final/initial state
interactions).  At each beam energy all events were spread over bins
corresponding to different Fermi momenta, ie. different energies in the cm.
frame. Usually theoretical predictions are compared with the energy dependence
of these points. However transparency depends on the energy of the particles in
the lab frame, rather than on the cm.\ energy. For fixed beam energy higher $s$
corresponds to higher Fermi momenta in the direction opposed to the incoming
proton.  Taking into account the results of the present paper, we expect
decreasing CT with increasing $s$ for fixed incident energy. This effect is
weak at 6 GeV/c due to the strong mixing of eigenstates, and the points do not
demonstrate strong energy dependence. At beam momentum of 12 GeV/c the expected
decrease of nuclear transparency is considerable, of the order of that
observed\cite{carroll}. Thus the points, that were considered as contradicting
the expected energy behavior now confirm the CT phenomenon more then other
points. There are some problems with the explanation of the data corresponding
to the beam momentum of 10 GeV/c. However they originate from only one point
which shows too high a transparency.

vii) As we stressed before the quasi elastic electron and proton scattering
have poor kinematics --- too low energy, $\nu$, of the recoil proton in
comparison with the square of the momentum transfer, $Q^2$. What processes are
better?  First, the quasielastic $A(p,2p)A'$ reaction is better with
asymmetrical geometry, with $Q^2$ and consequently the energy of the recoil
proton fixed.  The size of the recoil particle then does not depend on the
incident energy as well so all the incident energy dependence comes from the
time evolution of the proton wave function. Thus we can increase the incident
momentum without decreasing the cross section dramatically.

Acknowledgment: We would like to thank S.J.~Brodsky, O.~Ha\"usser,
W.B.~Lorenzon, G.A.~Miller and N.N.~Nikolaev for interest in the work and
useful discussions.  B.Z.K thanks TRIUMF for its hospitality during his visit
and B.K.J thanks the Natural Sciences and Engineering Research Council of
Canada for financial support.

\pagebreak

    Fig.~1. The nuclear transparency as a function of $x_B$. The curves
    correspond to $Q^2$ of 7 GeV$^2$ (long dashed curve), 15 GeV$^2$ (solid
    curve) and 30GeV$^2$ (short dashed curve). The dash-dotted curve is the
    Glauber model.

    Fig.~2. The nuclear transparency as a function of $Q^2$. The solid curve is
    for $x_B> 1$ while the dashed curve is for $x_B < 1$. The dash-dotted curve
    is the Glauber model.

\end{document}